\documentclass{article}

\usepackage{spconf}
\usepackage{amsmath,amsfonts,bm,amssymb,epsfig,psfrag,graphicx}
\usepackage{algorithm,algpseudocode}

\newtheorem{thm}{Theorem}

\newtheorem{defn}{Definition}
\newtheorem{prop}{Proposition}

\newcommand{\beq}{\begin{equation}}
\newcommand{\eeq}{\end{equation}}
\newcommand{\beqa}{\begin{eqnarray}}
\newcommand{\eeqa}{\end{eqnarray}}

\newcommand{\bnum}{\begin{enumerate}}
\newcommand{\enum}{\end{enumerate}}

\newcommand{\R}{\mathbb{R}}

\newcommand{\pf}{{\bf Proof: }}

\newcommand{\hspp}{\hspace{0.05in} }

\newcommand{\bfa}{ \mathbf{a} }

\newcommand{\Reps}{ R_{\epsilon} } 
\newcommand{\pr}{ \textrm{Pr} } 
\newcommand{\supp}{ {\mathrm{supp}} }

\newcommand{\magnf}{ {\mathrm{Mag}(\cdot)} }
\newcommand{\thetabm}{ {\boldsymbol{\theta}} }
\newcommand{\psibm}{ {\boldsymbol{\psi}} }
\newcommand{\thetahbm}{ \hat{\boldsymbol{\theta}} }
\newcommand{\deltabm}{ {\boldsymbol{\delta}} }

\newcommand{\omegabm}{ {\boldsymbol{\omega}} }

\newcommand{\Uth}{U_{\boldsymbol{\theta}}}
\newcommand{\snr}{ {\mathrm{SNR}} }
\newcommand{\snrf}{ {\mathrm{SNR}(\cdot)} }

\newcommand{\ADB}{adaptive distributed beamforming }

\title{A General Proof of Convergence for Adaptive Distributed Beamforming Schemes}

\name{Chang-Ching Chen, Chia-Shiang Tseng, and Che Lin
\thanks{This work is supported by National Science Council, R.O.C.,
under Grant NSC 99-2221-E-007-089-MY3.}}
\address{Institute of Communication Engineering \& Department of Electrical Engineering\\
National Tsing Hua University, \\ Hsinchu, Taiwan 30013 \\
\small E-mail: s9864511@m98.nthu.edu.tw,~s9964518@m99.nthu.edu.tw,~clin@ee.nthu.edu.tw}

\begin{document}
\ninept
\maketitle

\begin{abstract}
\vspace{-0.10cm}
This work focuses on the convergence analysis of adaptive distributed beamforming schemes that can be reformulated as local random search algorithms via a random search framework. Once reformulated as local random search algorithms, it is proved that under two sufficient conditions: a) the objective function of the algorithm is continuous and all its local maxima are global maxima, and b) the origin is an interior point within the range of the considered transformation of the random perturbation, the corresponding adaptive distributed beamforming schemes converge both in probability and in mean. This proof of convergence is general since it can be applied to analyze randomized adaptive distributed beamforming schemes with any type of objective functions and probability measures as long as both the sufficient conditions are satisfied. Further, this framework can be generalized to analyze an asynchronous scheme where distributed transmitters can only update their beamforming coefficients asynchronously. Simulation results are also provided to validate our analyses.
\end{abstract}


\vspace{-0.10cm}
\section{Introduction}\label{sec:intro}
\vspace{-0.20cm}
In a distributed network, distributed beamforming is considered as a promising scheme that allows distributed transmitters to convey common information efficiently in energy due to its potential array gain and low-complexity. However, to achieve distributed beamforming, or phase alignment, channel state information (CSI) at the transmitters is required and the cost of obtaining perfect CSI is too expensive in practice. Therefore, instead of obtaining perfect CSI, an adaptive scheme that uses a one-bit feedback link to acquire partial CSI at the transmitter ends was proposed in \cite{Mudumbai2006}, and the analyses of this one-bit feedback adaptive scheme have been extensively studied in \cite{RM07}\nocite{Johnson08, Thukral07, Bucklew08}-\cite{Lin10}.
In \cite{Johnson08}, \cite{Thukral07}, a discrete version of the one-bit adaptive scheme with binary signaling is considered. 
To analyze the behavior of the one-bit adaptive scheme, the authors in \cite{Bucklew08} applied stochastic approximations to show the convergence of the scheme. It has been shown that the sample path of the one-bit adaptive scheme approximately follows an ordinary differential equation under suitable conditions, and the convergence of the scheme is established accordingly. 
Alternatively, the one-bit adaptive scheme was reformulated as a local random search algorithm via a random search framework and analyzed in our previous work \cite{Lin10}. This reformulation allows us to use the techniques studied in the mature field of random search for systematically analyzing the characteristics of the one-bit adaptive scheme. With the help of this framework, it has been shown, without any approximation, that the one-bit adaptive scheme converges both in probability and in mean, and its convergence time scales linearly with the number of distributed transmitters. 

In this paper, we further generalize the framework in \cite{Lin10} and provide a more general proof of convergence and linear scalability for \ADB schemes. We show that any \ADB scheme that can be reformulated as a local random search algorithm converges both in probability and in mean if the algorithm satisfies two sufficient conditions for the objective function and the random perturbation. 
This framework is more general since once the sufficient conditions are satisfied, there is no need to specify the objective functions and probability measures. 
Instead of focusing on a particular objective function and symmetric probability measures as in \cite{Mudumbai2006}, \cite{Bucklew08}, \cite{Lin10}, our framework can be applied to analyze a much broader set of \ADB schemes and hence provide more potential to unify theoretical analysis in this field.
Furthermore, since the transmitters are deployed in a distributed fashion, it is possible that they experience different environments that cause their local clocks to be asynchronous. Our framework can be further extended to analyze \ADB schemes in such asynchronous setting by a straightforward application of Bayes' rule.

\vspace{-0.10cm}
\section{System Setup}\label{sec:syssetup}
\vspace{-0.20cm}
We consider a network with $N$ distributed transmitters that attempt to convey a common message to the receiver. We assume that all transmitters and the receiver are equipped with one antenna and the channels between each transmitter and the receiver are slow faded and frequency flat. Furthermore, we assume a noncoherent communication model where both the transmitters and the receiver do not have CSI. However, there is an error-free, zero-delay, and low-rate feedback link from the receiver to each transmitter so that the receiver can help aligning the beamforming phases through this reverse feedback link. The discrete-time, complex baseband received signal can be expressed as 
\beq \label{eq:model}
y[n] = \sum_{i=1}^{N}h_ig_i[n]s[n]+w[n] = \sum_{i=1}^{N}a_ib_i[n]e^{j(\phi_i+\psi_i[n])}s[n]+w[n]
\eeq
where $y[n]\in \mathbb{C}$ is the received signal, $s[n]\in \mathbb{C}$ is the common message with the average power constraint $E[|s[n]|^2]\le P_s$ for all $n$, and $w[n] \sim \mathcal{CN}(0,\sigma^2)$ is the additive white Gaussian noise. For transmitter $i$, we denote the time-invariant channel fading gain by $h_i=a_ie^{j\phi_i}\in \mathbb{C}$, and the beamforming coefficient by $g_i[n]=b_i[n]e^{j\psi_i[n]}\in \mathbb{C}$. 
For simplicity, we set $s[n] = \sqrt{P_s}$ and $b_i[n]=1$ and focus on the phase alignment during the training stage.

The SNR function at the receiver is given by 
\beq \label{eq:snr}
\snr(\thetabm[n])=\frac{P_s\left | \sum_{i=1}^{N} a_i e^{j\theta_i[n]} \right |^2}{\sigma^2}
\eeq
where $\thetabm[n]= \left[ \theta_1[n],\dots,\theta_{N}[n] \right]^T$ and $\theta_i[n]=\phi_i+\psi_i[n]$ is the total phase received at the receiver from transmitter $i$. 
The goal is to maximize the $\snrf$ so that the receiver can recover the signal with minimum error. However, since the receiver has neither the knowledge of CSI nor the objective function, \textit{i.e.}, the SNR function, it can only estimate a sample of the SNR function at each iteration. Therefore, this adaptive distributed beamforming problem can be reformulated as the problem stated as follows \cite{Lin10}:

\textit{Problem 1}: Given an unknown objective function $f:\Theta \to \R$, where $\Theta \subseteq \R^{N}$, and only samples of $f(\thetabm)$ are available for any $\thetabm \in \Theta$, find the global maxima of $f$.

This problem cannot be solved by gradient search method since there is no knowledge on the objective function and its gradient at the transmitter end in our case. However, from our previous work \cite{Lin10}, it can be solved by global random search algorithms in general. Further, if all the local maxima of the objective function are global maxima, local random search algorithms, which are more efficient, can also be applied.

\vspace{-0.10cm}
\section{General Proof of Convergence and Linear Scalability}\label{sec:pf}
\vspace{-0.20cm}
In this section, we provide a more general proof for the convergence and linear scalability of a set of adaptive distributed beamforming problems that can be reformulated as local random search algorithms. 
To this end, we briefly describe the local random search algorithm in \cite{Lin10} as follows:
\begin{itemize}

\item \emph{Step zero}: Initialize the algorithm by choosing $\thetabm[0] \in \Theta$.

\item \emph{Step one}: Generate a random perturbation $\deltabm[n]$ from the probability measure $\mu_n$ that could be time-varying and has the support $\supp(\mu_n)$. 

\item \emph{Step two}: Update the search point by $\thetabm[n] = D(\thetabm[n-1],\deltabm[n])$, where the mapping $D:\Theta \times \supp(\mu_n)  \to \Theta$ satisfies the condition that
\beq\nonumber
f(D(\thetabm[n-1],\deltabm[n])) \geq f(\thetabm[n-1]) 
\eeq

\end{itemize}
In most cases, the mapping $D$ can be expressed as
\begin{align}
&D(\thetabm[n-1], \deltabm[n])\nonumber \\ 
&=\thetabm[n-1]+G_n(\deltabm[n])1_{\left\{ f(\thetabm[n-1]+G_n(\deltabm[n])) > f(\thetabm[n-1])\right\}}\label{eq:mapping}
\end{align}
where $1_{\left\{\cdot\right\}}$ is the indicator function and $G_n:\supp(\mu_n) \to \Gamma_n\subseteq \R^N$ can be a general transformation. Note that $\Theta$ and $\Gamma_n$ are in the same space, $i.e.,\ \R^N$.

Now, referring to the local random search algorithm described above, we further provide two sufficient conditions for our proof.

\begin{enumerate}

\item[(S1)] The objective function $f$ is continuous and all its local maxima are global maxima. 
\item[(S2)] The origin is an interior point of the range of the transformation $G_n,\ i.e.,\ \Gamma_n$, for all $n$.

\end{enumerate}
In the following analyses, we consider local random search algorithms that satisfy both the sufficient conditions. 
Note that we do not need to specify the exact expressions of the objective function and the probability measure once the sufficient conditions are satisfied. This makes our proofs much more general than those in \cite{Lin10}.

\subsection{Convergence}
We first define the convergence in probability as follows.
\vspace{-0.10cm}
\begin{defn}
A sequence $\left\{\thetabm[n]\right\}_{n=0}^{\infty}$ generated by a random search algorithm is said to converge in probability if given $\epsilon>0$,
\beq\nonumber
\lim_{n \rightarrow \infty} \pr\left[ \thetabm[n] \in \Reps \right] = 1
\eeq
where
\beq\nonumber\label{eq:def_Reps}
\Reps := \left\{ \thetabm \in \Theta : f(\thetabm) > f\left(\thetabm^{*}\right) -\epsilon \right\}
\eeq
is the $\epsilon$-convergence region. In other words, $f(\thetabm[n])$ converges to $f(\thetabm^*)$ in probability, where $\thetabm^*$ is a global maximum point.
\end{defn}

Now, we further derive two propositions for the proof of convergence. For notational simplicity, we omit the time indices and write $\thetabm$, $\deltabm$ instead of $\thetabm[n]$, $\deltabm[n]$ in the following propositions and the corresponding proofs.

\vspace{-0.10cm}
\begin{prop} \label{prop:1}
Given an objective function $f$ and a transformation $G_n$ with domain $\supp(\mu_n)$ that satisfy the sufficient conditions (S1) and (S2). Let $D$ be the mapping as defined in \eqref{eq:mapping}, then for any $\thetabm \in \Theta \setminus \Reps$, there exists a set $A(\thetabm) \subset Range\{D(\thetabm, \cdot)\}$ with nonempty interior points, such that $f(\bfa)>f(\thetabm)$ for all $\bfa \in A(\thetabm)$, where $Range\{D(\thetabm,\cdot)\}$ is the range of the mapping $D$ at $\thetabm$.
\end{prop}

\vspace{-0.10cm}
\pf We begin the proof by defining a $f(\thetabm)$-superlevel set
\beq \nonumber
U_\thetabm := \left \{ \psibm \in \Theta : f(\psibm) \geq f(\thetabm) \right \}
\eeq
which is the subset of $\Theta$ with all the elements in $\Uth$ mapping to function values no less than $f(\thetabm)$. Note that the assumption $\thetabm \not \in \Reps$ implies that $\Reps \subseteq U_\thetabm$ and since $\Reps$ is nonempty, $\Uth$ is nonempty. Furthermore, since $f$ is continuous, the boundary points are those points that satisfy $f(\psibm)=f(\thetabm)$.

We claim that for any $\xi>0$, there exists a set $A(\thetabm) \subset B(\thetabm, \xi) \cap \Uth$ with nonempty interior points, such that $f(\bfa) > f(\thetabm)$ for all $\bfa \in A(\thetabm)$, where $B(\thetabm, \xi)$ is an open ball centered at $\thetabm$ with radius $\xi$. 
To this end, we first rule out the case where there are flat regions outside $\Reps$, that is, for some $\thetabm \not \in \Reps$, there exists a $\zeta>0$ such that $f(\psibm)=f(\thetabm)$ for all $\psibm \in B(\thetabm, \zeta)$. This case cannot happen since if it does, then $f(\psibm) \le f(\thetabm)$ for all $\psibm \in B(\thetabm, \zeta)$ and this implies that $\thetabm$ is a local maxima point and thus a global maxima point by the sufficient condition (S1), and this contradicts to the assumption that $\thetabm \not \in \Reps$. Therefore, there are no flat regions outside $\Reps$.

Excluding the existence of flat regions outside $\Reps$, we now consider two cases where $\thetabm$ is either a local minimum point or not. Note that the arguments leading to the absence of flat regions outside $\Reps$ implies that $\thetabm$ cannot be a local maxima point.

Case I:
If $\thetabm$ is a local minimum point, we can show that it is an interior point of $\Uth$ since the fact that $\thetabm$ is a local minimum point but not a local maxima point implies that there exists a $\xi>0$ such that $f(\psibm) > f(\thetabm)$ for all $\psibm \in B(\thetabm, \xi) \setminus\{\thetabm\}$. This means that $B(\thetabm, \xi) \subseteq \Uth$ and thus $\thetabm$ is an interior point of $\Uth$. Note that in this case, the entire $B(\thetabm, \xi)$ are in the interior of $\Uth$ and we choose $A(\thetabm)=B(\thetabm, \xi) \setminus\{\thetabm\}$, which obviously has nonempty interior points.

Case II:
If $\thetabm$ is not a local minimum point, it can be shown that it is a boundary point of $\Uth$ but not an isolated point. First, we show that $\thetabm$ is not an isolated point. If $\thetabm$ is an isolated point, it means that there exists a $\zeta>0$ such that for all $\psibm \in B(\thetabm, \zeta) \setminus \{\thetabm\}$, $\psibm \not \in \Uth$, or equivalently, $f(\psibm)<f(\thetabm)$. This means that $\thetabm$ is a local maximum point and thus a global maximum point by the sufficient condition (S1). Again, we obtain a contradiction to the assumption that $\thetabm \not \in \Reps$. Now, since $\thetabm$ is neither a local maximum point nor a local minimum point, then for any $\xi>0$, we can always find a point $\psibm_1 \in B(\thetabm,\xi)$ such that $f(\psibm_1) > f(\thetabm)$, i.e., $\psibm_1 \in \Uth$ and a point $\psibm_2 \in B(\thetabm,\xi)$ such that $f(\psibm_2) < f(\thetabm)$, i.e., $\psibm_2 \not \in U_\thetabm$. Thus, for any $\xi>0$, $B(\thetabm,\xi)$ contains both points inside and outside $\Uth$. This means that $\thetabm$ is a boundary point. Note that $\psibm_1$ is an interior point in $\Uth$ since if it is not, then it is a boundary point of $\Uth$ and it means $f(\psibm_1)=f(\thetabm)$, which contradicts to the statement that $f(\psibm_1) > f(\thetabm)$.
In this case, we choose $A(\thetabm)=B(\thetabm, \xi) \cap \Uth \setminus C$, where $C:=\{\psibm \in \Uth : f(\psibm) = f(\thetabm) \}$ and there are nonempty interior points in $A(\thetabm)$. Therefore, our claim is indeed true.

Now we define 
\beq\nonumber
\Omega := \left\{ \omegabm \in \Theta : \omegabm = \thetabm+G_n(\deltabm), \hspp \deltabm \in \supp(\mu_n) \right\}
\eeq
as the set by shifting $\Gamma_n$ with respect to $\thetabm$. Then the range of $D(\thetabm, \cdot)$ for any $\thetabm \not \in \Reps$ is given by
\beq\nonumber
Range\{ D\left ( \thetabm, \cdot \right ) \} = \Omega \cap U_\thetabm
\eeq
Note that the sufficient condition (S2) implies that $\thetabm$ is an interior point of $\Omega$. Since $\thetabm$ is also a boundary or interior point of $\Uth$, there always exist $\xi>0$ such that $B(\thetabm, \xi) \cap \Uth \subseteq Range\{ D\left ( \thetabm, \cdot \right ) \}$. Therefore, we can conclude that $A(\thetabm)\subset Range\{D(\thetabm, \cdot)\}$.$\hfill \blacksquare$

The following proposition states that for any $\thetabm$ outside $\Reps$, there is a non-zero probability to improve $f$ by applying a local perturbation to $\thetabm$. 

\begin{prop}\label{prop:pos_improve}
For any given $\thetabm \in \Theta \setminus \Reps$, there correspond $\gamma >0$ and $0< \eta \leq 1$ such that
\beq\nonumber
\pr \left[ f(\thetabm+G_n(\deltabm)) - f(\thetabm) \geq \gamma \right] \geq \eta
\eeq
where $\deltabm$ is a random vector generated with the probability measure $\mu_n$.
\end{prop}

\pf \textit{Proposition \ref{prop:1}} implies that for any $\thetabm \in \Theta \setminus \Reps$, there exists an interior point $\thetahbm \in A(\thetabm)$ and $\xi>0$ such that $\forall \psibm \in T:=B(\thetahbm, \xi)$, $f(\psibm)-f(\thetabm)\ge \gamma(\thetabm)$. Then, 
\beq\nonumber
\pr \left[ f(\thetabm+G_n(\deltabm)) - f(\thetabm) \geq \gamma(\thetabm) \right] \geq \mu_n (T) =: \eta(\thetabm) >0
\eeq
Note that $T$ is a function of $\thetabm$ since $\thetahbm$ depends on $\thetabm$. We complete the proof by letting 
\begin{align*}\nonumber
\gamma &=\inf_{\thetabm \in \Theta \setminus \Reps} \gamma(\thetabm) \\\nonumber
\eta &=\inf_{\thetabm \in \Theta \setminus \Reps} \eta(\thetabm)
\end{align*}
$\hfill \blacksquare$

Since there is always a non-zero probability to improve $f(\cdot)$ for each time step before the sequence reaches $R_{\epsilon}$, the convergence is expected. We describe this more precisely in the following theorem.

\begin{thm}\label{thm:converge}
For an objective function $f$ and a transformation $G_n$ with domain $\supp(\mu_n)$ that satisfy the sufficient conditions (S1) and (S2), let $\left\{\thetabm[n]\right\}_{n=0}^\infty$ be a sequence generated from the local random search algorithm. Then the resulting sequence converges in probability, i.e., given $\epsilon>0$,
\beq\nonumber
\lim_{n \rightarrow \infty} \pr\left[ \thetabm[n] \in \Reps \right] = 1
\eeq
\end{thm}

With \textit{Proposition \ref{prop:pos_improve}}, the proof directly follows the one provided in our previous work \cite{Lin10} by replacing the $\magnf$ in \cite{Lin10} by a general objective function $f(\cdot)$. We hence omit the details. Note that since $\left\{f(\thetabm[n])\right\}_{n=0}^\infty$ is non-negative and monotonically non-decreasing, this sequence also converges in mean by the Monotone Convergence Theorem \cite{Durrett95}.

\subsection{Linear Scalability}
For the analysis of the scaling law, we use an alternative definition of convergence. 
\begin{defn}
A sequence $\{\thetabm[n]\}_{n=0}^{\infty}$ generated by a random search algorithm is said to converge in mean if there exists an $M \geq 0$ such that 
\beq
E_{\{\deltabm[m]\}_{m=0}^{n} | \thetabm[0]}\left[f\left( \thetabm[n] \right)\right]>f\left( \thetabm^* \right) - \epsilon \label{eq:cvg_in_mean}
\eeq
for all $n\geq M$.
That is, $f(\thetabm[n])$ converges to $f(\thetabm^*)$ in mean. 
Furthermore, the iterations required to converge in mean is defined as the hitting time.
\end{defn}

With the above definition, we introduce the following theorem for linear scalability.

\begin{thm}
Given $\Theta \subseteq \mathbb{R}^N$, the \textbf{hitting time} of the random search algorithm scales linearly with $N$. That is, there exists $k < \infty$, such that Eqn. \eqref{eq:cvg_in_mean} holds for $M \le kN$.
\end{thm} 

By substituting the $\magnf$ in \cite{Lin10} by a more general objective function $f(\cdot)$ which satisfies the stated sufficient conditions, the proof directly follows the one provided in \cite{Lin10}.

Even though the derivation of the proofs for the above two theorems directly follow those in \cite{Lin10} with only little modification, we emphasize again that we do not need to specify the objective function and the probability measure used as in \cite{Lin10} once the sufficient conditions are satisfied. Therefore, our proofs are more general.

\vspace{-0.20cm}
\section{Asynchronous Scheme}\label{sec:asyn}
\vspace{-0.20cm}
In this section, we model an asynchronous \ADB scheme and analyze its convergence by extending our random search framework. 
We consider the case where at each time instance based on the global clock, only $\rho \%$ of the distributed transmitters update their phases and the rest of them keep their phases unchanged. 
This scenario can be equivalently modeled by assuming that each transmitter perturbs its phase independently with probability $p=\rho \%$ at each time slot. This is similar to but different from the $\rho \%$ algorithm proposed in \cite{Bucklew08} since in our case, we assume that all the transmitters transmit at all times but only on average $\rho \%$ of them change their phases at each time slot. 

Now, we show that this \ADB scheme still converges even when the phases are updated asynchronously. Similar to the analysis in Sec.~\ref{sec:pf}, we first derive a proposition stating that for each perturbation, there is a non-zero probability to improve the objective function $f$.

\vspace{-0.10cm}
\begin{prop}\label{prop:rho_improve}
Consider the asynchronous scheme described above, then for any given $\thetabm \in \Theta \setminus \Reps$, there corresponds $\gamma >0$ and $0<\lambda \leq 1$ such that
\beq\nonumber
\pr \left[ f(\thetabm+G_n(\deltabm)) - f(\thetabm) \geq \gamma \right] \ge \lambda
\eeq
where $\deltabm$ is a random vector generated with the probability measure $\mu_n$.
\end{prop}

\pf We prove the proposition by Bayes' rule. Let $Z$ be the event that all transmitters update their phases. Then
\beqa
&\ &\pr \left[ f(\thetabm+G_n(\deltabm)) - f(\thetabm) \geq \gamma \right]  \nonumber\\ 
&\ge& \pr \left[ f(\thetabm+G_n(\deltabm)) - f(\thetabm) \geq \gamma \mid Z \right] \pr \left[ Z \right] \nonumber \\
&\ge& \eta p^{N} =:\lambda \nonumber
\eeqa
The last inequality follows from \textit{Proposition \ref{prop:pos_improve}}. $\hfill \blacksquare$

With \textit{Proposition \ref{prop:rho_improve}}, we can apply \textit{Theorem \ref{thm:converge}} to show that this asynchronous scheme converges both in probability and in mean. Note that by applying the simple Bayes' rule, we are able to prove the convergence for a much broader set of \ADB schemes. This demonstrates again the generality and flexibility of our framework.


\vspace{-0.20cm}
\section{Simulation Results and Conclusion}\label{sec:sim}
\vspace{-0.20cm}
In this section, numerical results are provided to validate our analyses. We consider a network with $N=200$ distributed transmitters and assume the channel coefficients are i.i.d. $\mathcal{CN}(0,1)$. Unless otherwise specified, we use the SNR function described by \eqref{eq:snr} as our objective function, set $G_n(\cdot)$ to be an identity function, and use a random perturbation with the time-invariant probability measure $\mu \sim \mathcal{U}[-\Delta_0, \Delta_0]^{N}$, where $\mathcal{U}[\cdot, \cdot]$ denotes the uniform distribution and $\Delta_0 = 5^{\circ}$ in the simulation. Besides, for simplicity, we say that the algorithm converges once $\snr(\thetabm) \ge \alpha \snr (\thetabm^*)$, where $\alpha=0.9$. It has been shown in~\cite{Lin10} that the SNR function, the probability measure $\mu$, and the transformation $G_n(\cdot)$ considered above satisfy both the sufficient conditions. We hence use this setting to verify our analyses.


\begin{figure}[tb]
\psfrag{No. of Iterations}[ct][lc][0.8]{No. of Iterations}
\psfrag{Proportion to Max}[lb][lc][0.8]{Proportion to Max}
\psfrag{(a)}[cb][lc][0.8]{(a)}
\psfrag{(b)}[cb][lc][0.8]{(b)}
\psfrag{title}{}
\centering{
\includegraphics[width = 3.6 in, height= 2.1 in ]{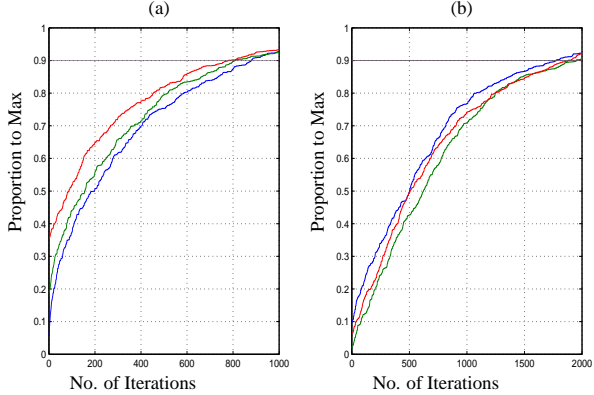} 
\caption{The convergence of (a) the objective function $f$ defined in Sec. \ref{sec:sim} satisfying the sufficient condition (S1) and (b) the asymmetric probability measure satisfying the sufficient condition (S2)}
\label{fig:cvg}
}
\end{figure}

\begin{figure}[tb]
\psfrag{No. of Iterations}[lc][lc][0.8]{Average No. of Iterations}
\psfrag{rho}[lc][lc][0.7]{$\rho \%$}
\psfrag{(a)}[cb][lc][0.8]{(a)}
\psfrag{(b)}[cb][lc][0.8]{(b)}
\psfrag{Number of Distributed Transmitters}[lc][lc][0.7]{No. of Distributed Transmitters}
\centering{
\includegraphics[width= 3.6 in, height=2.1 in]{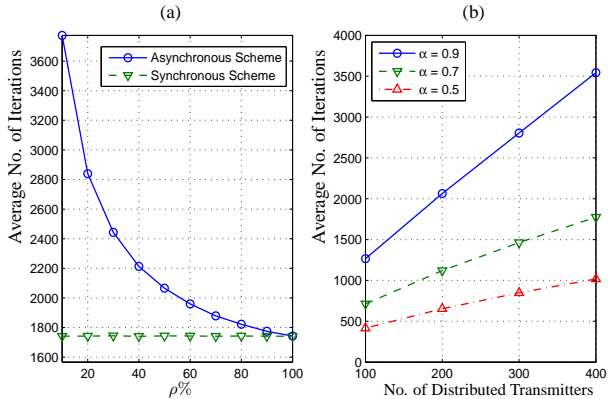} 
\caption{(a) The average iterations necessary for the convergence of the asynchronous scheme at different $\rho \%$. (b) The average convergence time of the asynchronous scheme with different $\alpha$}
\label{fig:asyn}
}
\end{figure}
Fig.~\ref{fig:cvg} shows the convergence of (a) an arbitrary objective function satisfying the sufficient condition (S1) and (b) a random perturbation with probability measure which is not symmetric but satisfies the sufficient condition (S2). For Fig.~\ref{fig:cvg}(a), we choose the objective function to be $f(\thetabm)=\sum_{i=1}^{N} \left [ -\left ( (\theta_i\ \text{mod}\ \pi) - \frac{\pi}{2} \right)^2 + (\frac{\pi}{2})^2 \right ]$, where mod is the modulo function.  
It can be shown that this function satisfies the sufficient condition (S1) and we omit the details due to space constraint. 
For Fig.~\ref{fig:cvg}(b), we generate the asymmetric distribution by randomly shifting $\mu$ around zero. For both settings, we demonstrate that the modified algorithm either with new objective function or asymmetric probability measure converges from different initial points. Note that more simulations with different random shifts of $\mu$ have been done and show similar behaviors but are not included here due to space constraint. The asynchronous scheme described in Sec.~\ref{sec:asyn} is considered in Fig.~\ref{fig:asyn}. In Fig.~\ref{fig:asyn}(a), we show the difference between the asynchronous scheme with different $\rho \%$ and the synchronous scheme in terms of the convergence time. This figure also shows the average iterations necessary for the asynchronous scheme to converge for different $\rho \%$. It is obvious that the performance of the asynchronous scheme is worse than that of the synchronous scheme. However, we emphasize that even when only $\rho \%$ of the distributed transmitters update their phases at a given time instance, the algorithm still converges. Since the convergence behavior from different initial points of the asynchronous scheme for any $\rho \%$ is also similar to those in Fig.~\ref{fig:cvg}, we omit it due to space limitation. In Fig.~\ref{fig:asyn}(b), we demonstrate the linear scalability of the asynchronous scheme with $\rho\%=50\%$. We observe that the convergence time increases with $\alpha$, and that for any $\alpha$, the average number of iterations necessary for the convergence scales linearly with the number of distributed transmitters. These simulation results verify our theoretical analyses.


In this paper, we generalized the proof of convergence for adaptive distributed beamforming schemes that can be reformulated as local random search algorithms and 
satisfy two sufficient conditions. Specifically, we have shown that such adaptive distributed beamforming schemes converge both in probability and in mean and their convergence time scales linearly with the number of distributed transmitters. We further extended our framework and analyzed the case where distributed transmitters can only update their beamforming coefficients asynchronously. Simulations were provided to validate our analyses.


\bibliographystyle{IEEEtran}
\footnotesize
\bibliography{ICASSP2011_Ref.bib}

\end{document}